%% file: LcinPbPb_paper_final.tex
\documentclass[ALICE,manyauthors]{cernphprep}
\pdfoutput=1
\usepackage{hyperref}
\usepackage{lineno}
\usepackage{xcolor}

\usepackage[english]{babel}
\usepackage{graphicx}
\usepackage{verbatim}
\usepackage{amsfonts}
\usepackage{makeidx}
\usepackage{color}
\usepackage{amsmath}
\usepackage{lineno}
\usepackage{epsfig}
\usepackage{epstopdf}
\usepackage{hyperref}
\usepackage{cite}
\usepackage{textcomp}
\newcommand {\pp}        {\mbox{$\mathrm {p\kern-0.05em p}$}\xspace}
\mathchardef\mhyphen="2D 
\newcommand{\fdwn}       {\mathrm{feed}\mhyphen\mathrm{down}}

\newcommand{\sqrts}{\sqrt{s}}
\newcommand{\sqrtsNN}{\sqrt{s_{\rm \scriptscriptstyle NN}}}

\newcommand{\TeV}{\ensuremath{\mathrm{TeV}}\xspace}
\newcommand{\tev}{\ensuremath{\mathrm{TeV}}\xspace}
\newcommand{\gevc}{{\rm GeV}/\ensuremath{c}\xspace}

\newcommand{\pt}{\ensuremath{p_{\rm T}}\xspace}
\newcommand{\pT}{\pt}

\newcommand{\mub}{\mathrm{\mu b}}

\newcommand{\AuAu}{\ensuremath{\mbox{Au--Au}}\xspace}
\newcommand{\PbPb}{\ensuremath{\mbox{Pb--Pb}}\xspace}
\newcommand{\pPb}{\ensuremath{\mbox{p--Pb}}\xspace}
\newcommand{\KzStopippim}{\ensuremath{\rm K^{0}_{S}\to \pi^{+} \pi^{-}}\xspace}

\newcommand{\Raa}{R_{\rm AA}}

\newcommand{\RAA}{R_{\rm AA}}
\newcommand{\TAA}{T_{\rm AA}}

\newcommand{\DtoKpi}{{\rm D}^0 \to {\rm K}^-\pi^+}

\newcommand{\Dzero}{\ensuremath{\rm D^0}\xspace}

\newcommand{\Ds}{{\rm D_s^+}}

\newcommand{\dEdx}{{\rm d}E/{\rm d}x}

\newcommand{\Lc}{\mbox{$\mathrm {\Lambda_{c}}$}\xspace}
\newcommand{\Lcplus}{\ensuremath{\rm \Lambda_c^+}}
\newcommand{\AntiLc}{\ensuremath{\rm  {\overline{\Lambda}{}_c ^-\xspace}}}
\newcommand{\LctopKzeros}{\rm \Lambda_c^+ \to p K_S^0}

\newcommand{\LcOverDzero}{\Lcplus/\Dzero}
\newcommand{\Kzeros}{\rm K_S^0}
\newcommand{\Nraw}{N_{\mathrm{raw}}}
\newcommand{\fprompt}{f_{\mathrm{prompt}}}

\begin{document}
\begin{titlepage}
\PHyear{2018}
\PHnumber{261}
\PHdate{27 September}  
\title{$\Lambda_{\mathrm{c}}^{+}$ production in Pb--Pb collisions at $\sqrtsNN$ = 5.02~TeV}
\ShortTitle{$\Lambda_{\mathrm{c}}^{+}$ production in Pb--Pb collisions at $\sqrtsNN=5.02$~TeV}   
\Collaboration{ALICE Collaboration\thanks{See Appendix~\ref{app:collab} for the list of collaboration members}}
\ShortAuthor{ALICE Collaboration} 
\begin{abstract}
A measurement of the production of prompt $\Lcplus$ baryons in Pb--Pb collisions at $\sqrtsNN = 5.02$~TeV with the ALICE detector at the LHC is reported. The $\Lcplus$ and $\AntiLc$ were reconstructed at midrapidity ($|y| < 0.5$) via the hadronic decay channel $\LctopKzeros$ (and charge conjugate) in the transverse momentum and centrality intervals $6 < \pt <12$~\gevc and 0--80\%.
The $\Lcplus/\Dzero$ ratio, which is sensitive to the charm quark hadronisation mechanisms in the medium, is measured and found to be larger than the ratio measured in minimum-bias \pp collisions at $\sqrts = 7$~TeV and in \pPb collisions at $\sqrtsNN = 5.02$~TeV. In particular, the values in \pPb and \PbPb collisions differ by
about two standard deviations of the combined statistical and systematic uncertainties in the common $\pt$ interval covered by the measurements in the two collision systems. 
The $\Lcplus/\Dzero$ ratio is also compared with model calculations including different implementations of charm quark hadronisation. 
The measured ratio is reproduced by models implementing a pure coalescence scenario, while adding a fragmentation contribution leads to an underestimation.
The $\Lcplus$ nuclear modification factor, $\RAA$, is also presented. The measured values of the $\RAA$ of $\Lcplus$, $\Ds$ and non-strange D mesons are compatible within the combined statistical and systematic uncertainties. They show, however, a hint of a hierarchy $(\RAA^{\Dzero}<\RAA^{\Ds} <\RAA^{\Lcplus})$, conceivable with a contribution from coalescence mechanisms to charm hadron formation in the medium.

\end{abstract}
\end{titlepage}
\setcounter{page}{2}

\section{Introduction}
\label{sec:Introduction}

Measurements of the production of open-heavy flavour hadrons in heavy-ion collisions provide important information on the properties of the Quark\textendash Gluon Plasma (QGP), the state of strongly-interacting matter formed at the very high temperatures and energy densities reached in heavy-ion collisions~\cite{Braun-Munzinger:2015hba, PhysRevD.90.094503}.
Several measurements of the production and elliptic flow of D mesons and leptons from the decay of heavy-flavour hadrons in \PbPb collisions at the LHC and in \AuAu collisions at RHIC~\cite{Andronic:2015wma,Prino:2016cni} indicate that charm quarks interact strongly with the medium constituents. In-medium energy loss is studied via the nuclear modification factor, $\RAA$, defined as the ratio of the yield in \PbPb collisions and that in \pp collisions scaled by the number of binary nucleon\textendash nucleon collisions.
A model~\cite{Greco:2003vf, Oh:2009zj} including a significant fraction of low and intermediate transverse momentum (\pT) charm and beauty quarks hadronising via coalescence (or recombination) with light quarks from the medium better describes the experimental results.
This mechanism is expected to also affect the production of $\Ds$ given the strange-quark rich environment of the created medium. At higher transverse momentum (\pT$>7$~\gevc at LHC energies~\cite{Plumari:2017ntm}) hadronisation by vacuum fragmentation is expected to be the dominant production mechanism. 

In this context, the study of charm baryons is essential to understand charm hadronisation.
Models including coalescence predict an enhanced baryon-to-meson ratio at low and intermediate transverse momentum in comparison to that expected in \pp collisions. This effect adds to the hadron-mass dependent transverse-momentum shift due to the presence of radial flow in heavy-ion collisions, that is able to explain the observed increase of the baryon-to-meson ratio in the light sector up to about 2 \gevc~\cite{Abelev:2013xaa}. The study of non-strange D-mesons, $\Ds$ and \Lcplus~could help to disentangle the role of coalescence and radial flow, because of the smaller mass differences than for light-flavour hadrons.

For the particular case of charm baryons, the possible existence of light di-quark bound states in the QGP could further enhance the \Lcplus/\Dzero ratio in the coalescence model~\cite{Lee:2007wr}. An enhancement of the \pT-integrated \Lcplus/\Dzero ratio in the presence of a QGP is also predicted by the statistical hadronisation model~\cite{Kuznetsova:2006bh}, where at LHC energies the relative abundance of hadrons depends on their masses, their flavour content and the freeze-out temperature of the medium. 
In addition, an enhancement of charm-baryon production in \PbPb collisions would make the charm baryons an important fraction of the total charm production cross section. 

The study of a potential enhancement effect in charm-baryon production in relativistic heavy-ion collisions requires a baseline reference in smaller collision systems. The \Lcplus-baryon production was measured by the ALICE Collaboration in \pp collisions at $\sqrts = 7$~\TeV in the transverse momentum and rapidity ($y$) intervals $1<\pT<8$~\gevc and $|y| < 0.5$~\cite{Acharya:2017kfy}. 
The obtained baryon-to-meson ratio is larger than previous measurements at lower centre-of-mass energies and in different collision systems (see Ref.~\cite{Acharya:2017kfy} and references therein), and also higher than the results reported by the LHCb Collaboration in \pp collisions at $\sqrts = 7$~\tev~in the rapidity range $2.0<y<4.5$~\cite{Aaij:2013mga}. Expectations from perturbative Quantum Chromodynamics (pQCD) calculations and Monte Carlo event generators underpredict the data, indica\-ting that the fragmentation of charm quarks is not fully understood~\cite{Acharya:2017kfy} and partially challenged by data collected so far at the LHC, as discussed extensively in Ref.~\cite{Maciula:2018iuh}.
The production of \Lcplus~baryons was also measured by the ALICE Collaboration in \pPb collisions at $\sqrtsNN = 5.02$~\tev in $2<\pT<12$~\gevc and $-0.96 < y < 0.04$~\cite{Acharya:2017kfy}, and a measurement in the same collision system by the LHCb Collaboration~\cite{Aaij:2018iyy} is also available. 
The \Lcplus~nuclear modification factor $R_{\rm pPb}$
is compatible with unity within statistical and systematic uncertainties. The baryon-to-meson ratios \Lcplus/\Dzero measured in \pp and \pPb collisions are compatible within uncertainties. A model~\cite{Li:2017zuj,Song:2018tpv} including hadronisation via coalescence in these collision systems has been proposed to describe the measurements at LHC energies.

This letter reports measurements of the production of the prompt charm baryon \Lcplus~and its charge conjugate in \PbPb collisions at $\sqrtsNN = 5.02$~\TeV with the ALICE detector~\cite{Aamodt:2008zz} at the LHC. Hereafter, $\Lambda_{\rm c}$ refers indistinctly to both particle and anti-particle, and all mentioned decay channels refer also to their charge conjugates. The \Lcplus corrected yield is obtained as the average of the particle and the anti-particle yield. The notation \Lcplus is used when referring to this average, and thus to indicate physics quantities such as the \Lcplus/\Dzero ratio.
The measurement was performed in the 0--80\% centrality class in the transverse momentum and rapidity intervals $6<\pT<12$ \gevc and $|y| < 0.5$.
Only prompt \Lc-baryons were considered: the beauty-hadron feed-down was subtracted, as described in the next section. The \Dzero-meson yield was obtained in the same transverse momentum and centrality interval as the \Lc-baryon, following the analysis procedure described in Ref.~\cite{Acharya:2018hre}.

 \section{Data sample and analysis strategy}
The measurement of the $\Lc$-baryon production was performed by reconstructing the decays $\LctopKzeros$ with a branching ratio (BR) equal to $(1.58 \pm 0.08)\%$ and \KzStopippim with BR = (69.20 $\pm$ 0.05)$\%$~\cite{Tanabashi:2018pdg}. The $\Dzero$ mesons were reconstructed in the decay channel $\DtoKpi$ with $\mathrm{BR} = (3.93 \pm 0.04)\%$~\cite{Tanabashi:2018pdg}. The $\Lc$ and $\Dzero$ candidates were reconstructed in the same transverse momentum, rapidity and centrality intervals. 
The analysis benefits from the tracking and particle identification capabilities of the ALICE central barrel detectors located within a large solenoidal magnet that provides a magnetic field of 0.5~T parallel to the LHC beam axis.
A complete description of the ALICE apparatus and its performance can be found in Refs.~\cite{Aamodt:2008zz, Abelev:2014ffa}. The main detectors used in this analysis include the Inner Tracking System (ITS)~\cite{Aamodt:2010aa}, the Time Projection Chamber (TPC)~\cite{Alme:2010ke}, the Time-Of-Flight detector (TOF)~\cite{Akindinov:2013tea} and the V0 detector~\cite{Abbas:2013taa} located inside the solenoidal magnet, as well as the Zero Degree Calorimeters (ZDC)~\cite{Aamodt:2008zz} located in the LHC tunnel at about $\pm112.5$~m from the nominal interaction point and composed of two proton and two neutron calorimeters.

The analysed data sample consists of about $83\times 10^{6}$ Pb--Pb collisions at $\sqrtsNN=5.02$~TeV, corresponding to an integrated luminosity of $\mathcal{L}_{\mathrm{int}}\approx 13.4~\mub^{-1}$. The interaction trigger was provided by the coincident signals from the two arrays of the V0 detector, 
covering the pseudorapidity intervals $-3.7 < \eta < -1.7$ and $2.8 < \eta < 5.1$. 
Background events from beam\textendash gas interactions were removed in the offline analysis using the timing information provided by the V0 and the neutron ZDC. Only events with a primary vertex reconstructed within $\pm 10$~cm from the centre of the detector along the beam line were considered for the analysis. Events were selected in the centrality class 0--80\%, defined in terms of percentiles of the hadronic Pb--Pb cross section, using the amplitudes of the signals in the V0 arrays~\cite{ALICE-PUBLIC-2018-011}. 
 
The $\Lc$ candidates were constructed by combining a proton candidate track with a $\Kzeros$ candidate identified through its V-shaped neutral decay topology (${\rm V^0}$). The charged tracks and the $\Kzeros$ candidates were selected as described in Ref.~\cite{Acharya:2017kfy} for pp collisions with additional requirements to reduce the larger combinatorial background due to the higher charged-track multiplicity in Pb--Pb with respect to pp collisions. In particular, candidate proton tracks were required to have a hit in the innermost ITS layer and tighter selections on the $\Kzeros$ were applied: a maximum distance of closest approach between the ${\rm V^0}$ decay tracks of 0.4~cm, a minimum cosine of the ${\rm V^0}$ pointing angle to the primary vertex of 0.9998, a minimum $\pt$ of the $\Kzeros$ candidates of 1~\gevc, and a cut in the Armenteros-Podolanski space~\cite{Armenteros} to remove contributions from $\Lambda$ decays.
The identification of protons was based on the specific ionisation energy loss $\dEdx$ in the TPC and on the time of flight measured with the TOF detector, using as a discriminating variable ($n_{\sigma}$) the difference between the measured value and the expected value for the proton mass hypothesis divided by the detector resolution.  A $|n_{\sigma}| < 3$ selection was applied on the TPC $\dEdx$ and TOF time-of-flight measurements for tracks with $\pt < 3$~\gevc. For tracks with $\pt > 3$~\gevc an asymmetric selection was used to limit the contamination from pions in the TPC and from kaons in the TOF and the requirements were $-3 < n_{\sigma}^{\mathrm{TPC}} < 2$ and $-2 < n_{\sigma}^{\mathrm{TOF}} < 3$ for the TPC and TOF signals. Tracks without TOF information were discarded. 
The $\Lc$ candidates were selected requiring a cosine of the proton emission angle in the $\Lc$ centre-of-mass system with respect to the $\Lc$ momentum direction larger than 0.5. A selection on the signed transverse impact parameter of the proton, i.e. the distance of closest approach between the proton track and the primary vertex, larger than 0.003~cm was also applied (the sign of the impact parameter is defined as positive when the angle between the $\Lc$ flight line and the momentum vector is smaller than $90^\circ$).

The $\Dzero$ candidates were reconstructed by combining pairs of tracks with the proper charge sign combination and selected in the interval $6 < \pt < 12$~\gevc   using the same criteria   described in Ref.~\cite{Acharya:2018hre} for the interval $6 < \pt < 7$~\gevc in the 10\% most central Pb--Pb collisions. 

After all selections, the acceptance in rapidity for $\Lc$ and $\Dzero$ candidates drops steeply to zero for $|y|>0.8$  in the $\pt$ interval used for the analysis. Therefore, a fiducial acceptance cut $|y|<0.8$  was applied as described in Refs.~\cite{Acharya:2017kfy} and ~\cite{Acharya:2018hre}. 

The $\Lc$ and $\Dzero$ raw yields were extracted by fitting the invariant mass distributions of the candidates passing the selection criteria. The fit functions consist of a Gaussian to describe the signal and an exponential to describe the background. In the case of the $\Lc$, the width of the Gaussian was fixed to the value obtained from Monte Carlo simulations. The stability of the $\Lc$ signal extraction was verified by fitting the invariant mass distribution after the subtraction of the background evaluated with an event-mixing technique and no discrepancy between the two approaches was observed. For the $\Dzero$-meson yield, the contribution of signal candidates with the wrong K--$\pi$ mass assignment (reflections) to the invariant-mass distribution was taken into account by including an additional term, parameterised from simulations with a double-Gaussian shape, in the fit function~\cite{Abelev:2014ipa}.

\begin{figure}[!t]
\begin{center}
\includegraphics[width=1.0\textwidth]{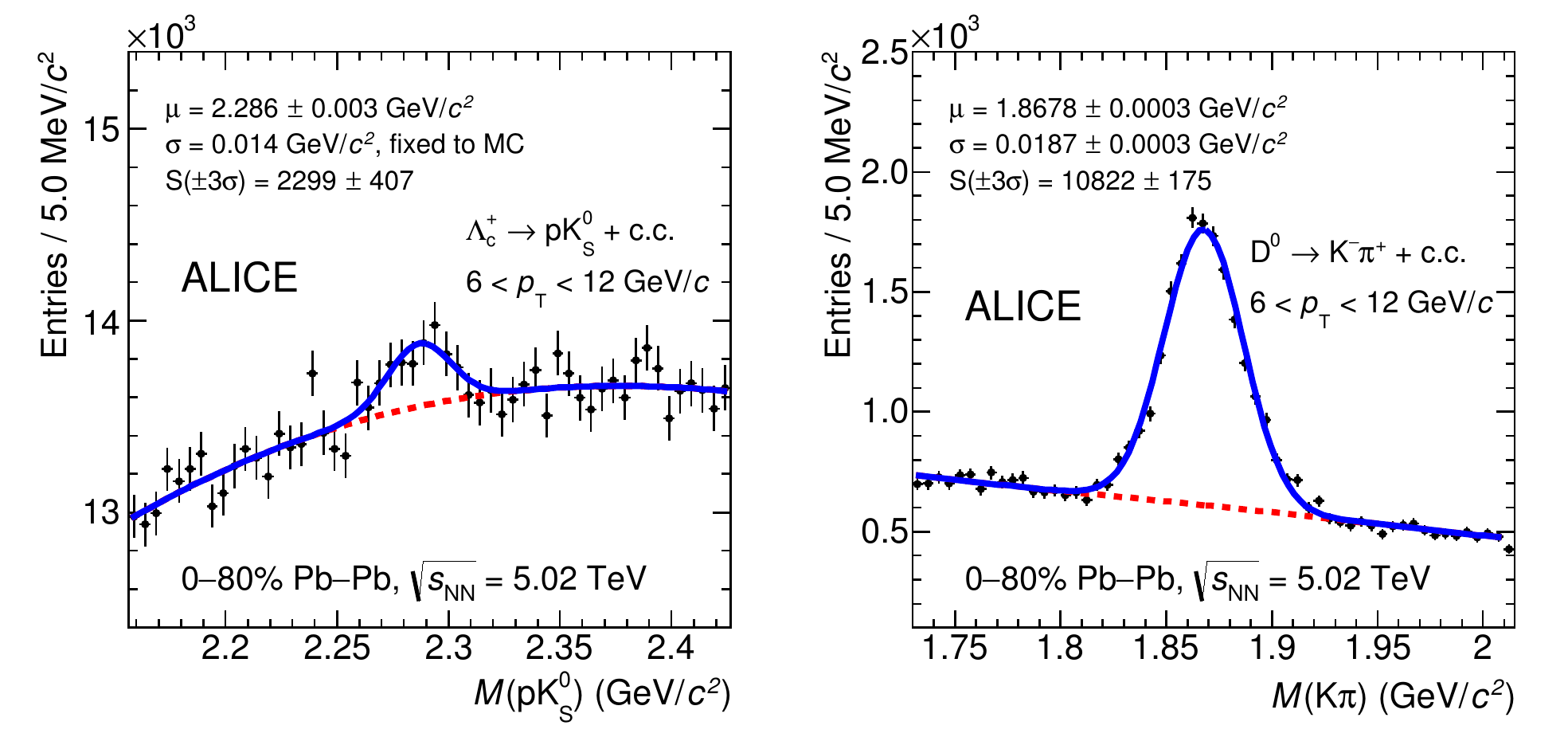}      
\caption{Invariant-mass distributions for the $\Lc$ (left) and $\Dzero$ (right) candidates in the momentum interval $ 6 < \pt< 12$~\gevc and for the 0--80\% centrality class. The dashed curves represent the fit to the background, while the solid curves represent the total fit function.}
\label{fig:InvMass} 
\end{center}
\end{figure}

The invariant mass distributions of the selected $\Lc$ and $\Dzero$ candidates are shown in Fig.~\ref{fig:InvMass}. 

The prompt $\Lcplus$ ($\Dzero$) production yield was calculated as

\begin{equation}\label{eq:yielddef}
\left. \frac{\mathrm{d}N_{\mathrm{prompt}}^{\Lcplus (\Dzero)}}{\mathrm{d}p_{\mathrm{T}}} \right|_{|y|<0.5} = \frac{1}{2}\frac{1}{c_{\Delta y}}\frac{1}{\Delta \pt} \frac{f_{\mathrm{prompt}} \cdot N_{\mathrm{raw}}|_{|y|<0.8}}{(\mathrm{Acc} \times \epsilon)_{\mathrm{prompt}} \cdot \mathrm{BR} \cdot N_{\mathrm{evt}}},
\end{equation}
where $\Nraw$ is the raw yield (sum of particles and anti-particles) in the transverse momentum interval of width $\Delta \pt$,  $f_{\mathrm{prompt}}$ is the fraction of prompt $\Lambda_{\mathrm{c}}$ ($\Dzero$) in the raw yield, $(\mathrm{Acc} \times \epsilon)$ is the product of acceptance and reconstruction efficiency for prompt $\Lambda_{\mathrm{c}}$ ($\Dzero$), BR is the branching ratio of the considered decay mode and $N_{\mathrm{evt}}$ is the number of events considered for the analysis. The correction factor for the rapidity coverage $c_{\Delta y}$ was computed as the ratio of the generated $\Lambda_{\mathrm{c}}$ ($\Dzero$) yield in $|y|<0.8$ and that in $|y|< 0.5$. The factor $1/2$ takes into account that the raw yield is the sum of particles and anti-particles, while the production yield is reported as their average.

The correction for the detector acceptance and reconstruction efficiency was determined by means of Monte Carlo (MC) simulations. The underlying Pb--Pb events at $\sqrtsNN = 5.02$~TeV were simulated using the HIJING v1.383~\cite{Wang:1991hta} generator and prompt and feed-down
$\Lc$ ($\Dzero$) were added using the PYTHIA v6.421~\cite{Sjostrand:2006za} generator with Perugia 11 tune. The generated particles were transported through the ALICE detector using the GEANT3~\cite{Brun:1994aa} package. A realistic detector response was introduced in the simulations to reproduce the performance of the ALICE detector system during data taking.

The $\pt$ distributions of the $\Lc$ and $\Dzero$ in PYTHIA were corrected in order to obtain more realistic distributions. 
The same \pt-dependent weighting factor, calculated as the ratio of the measured $\Dzero$ $\pt$ distribution in finer $\pt$ bins~\cite{Acharya:2018hre} and the one simulated with PYTHIA, was used for both particles.
The $\Lc$ and $\Dzero$ reconstruction efficiency in the large centrality class 0--80\% was obtained as the weighted average of the efficiencies in smaller centrality classes to take into account the variation of the efficiency and the scaling of the yields of the $\Lc$ baryons and $\Dzero$ mesons with centrality. The applied weights were calculated as the product of the $\RAA$ of the $\Dzero$ and the average number of nucleon\textendash nucleon collisions ($<N_{\mathrm{coll}}>$) in the centrality class considered~\cite{Acharya:2018hre}.
The $(\mathrm{Acc} \times \epsilon)$ value is about 6\% for prompt and about 9\% for feed-down $\Lc$ and about 8\% for prompt and about 11\% for feed-down $\Dzero$.

The prompt $\Lc$ ($\Dzero$) fraction, $\fprompt$, was calculated as

\begin{align}
  \fprompt &= 1-\left(\frac{N_{\fdwn}^{\Lambda_{\rm {c}} (\Dzero)}}{N_{\mathrm{prompt}}^{\Lambda_{\rm {c}} (\Dzero)}} \right) = \nonumber \\
  &= 1-\left< \TAA \right> \cdot \left.\frac{\mathrm{d}^2 \sigma}{\mathrm{d}y \mathrm{d}\pt}\right|^{\mathrm{FONLL}}_{\fdwn} \cdot R_{\mathrm{AA}}^{\fdwn} \cdot \frac{(\mathrm{Acc} \times \epsilon)_{\fdwn}\cdot c_{\Delta y} \cdot \Delta \pt \cdot \mathrm{BR} \cdot N_{\mathrm{evt}}}{\Nraw / 2}.
\end{align}
  
The contribution of $\Lc$ ($\Dzero$) from beauty-hadron decays was estimated using the FONLL~\cite{Cacciari:1998it, Cacciari:2001td} beauty-production cross sections as described in detail in Ref.~\cite{ALICE:2011aa}. 
The fraction of beauty quarks that fragment to beauty hadrons and subsequently decay into $\Lc$ baryons $f(\mathrm{b} \to \Lc) = 0.073$ was taken from Ref.~\cite{Gladilin:2014tba}. The beauty-hadron decay kinematics were modeled using the EVTGEN~\cite{Lange:2001uf} package. The $(\mathrm{Acc} \times \epsilon)_{\fdwn}$ term for both particles was calculated from the Monte Carlo simulations described above. The average nuclear overlap function, $\left< \TAA \right> $, was estimated via Glauber model calculations~\cite{Miller:2007ri,Loizides:2017ack}. 
In this formalism the nuclear modification factor $\RAA$ is then the ratio of the yield in \PbPb collisions and the production cross section in \pp collisions scaled by $\left< \TAA \right> $.

A hypothesis on the $\RAA^{\mathrm{\fdwn}}$ of feed-down $\Lc$ and $\Dzero$ is used. 
For the $\Dzero$, the hypothesis is the same as in other analyses (e.g. in Ref.~\cite{Acharya:2018hre}): the central value is obtained by assuming $\RAA^{\mathrm{\fdwn~\Dzero}}/\RAA^{\mathrm{prompt~\Dzero}} = 2$, justified by the CMS measurement of J/$\psi$ from B-meson decays~\cite{Khachatryan:2016ypw} and by the ALICE and CMS measurements of D mesons ~\cite{Acharya:2018hre,Sirunyan:2017xss} indicating that prompt charm mesons are more suppressed than non-prompt charm mesons. The ratio is varied in the interval $1< \RAA^{\mathrm{\fdwn~\Dzero}} / \RAA^{\mathrm{prompt~\Dzero}} <3$ to estimate the systematic uncertainty.
Since no measurements of beauty-baryon production in nucleus\textendash nucleus collisions are available, for the $\Lc$
the central hypothesis was taken from model calculations which predict $\RAA^{\mathrm{\fdwn~\Lcplus}}/\RAA^{\mathrm{prompt~\Lcplus}}=2$ when considering c and b quark fragmentation and energy loss in the medium~\cite{Das:2016llg}. 
The ratio $\RAA^{\mathrm{\fdwn~\Lcplus}}/\RAA^{\mathrm{prompt~\Lcplus}}$  was decomposed into two terms to estimate the uncertainty on the assumption:

\begin{equation}\label{eq:raa}
\frac{\RAA^{\fdwn~\Lcplus}}{\RAA^{\mathrm{prompt~\Lcplus}}} = \frac{\RAA^{\fdwn~\Dzero}}{\RAA^{\mathrm{prompt~\Dzero}}}\cdot\frac{\frac{(\LcOverDzero)_{\mathrm{PbPb,}\fdwn}}{(\LcOverDzero)_{\mathrm{pp,} \fdwn}}}{\frac{(\LcOverDzero)_{\mathrm{PbPb, prompt}}}{(\LcOverDzero)_{\mathrm{pp, prompt}}}}.
\end{equation}
 
The first term is the same as for the $\Dzero$ and thus the same hypothesis is adopted. 
The second term is varied in the range 0.5--1.5 to calculate the systematic uncertainty.
The upper limit is determined a-posteriori such that $\RAA^{\mathrm{\fdwn~\Lcplus}} < 2$ as suggested by the fact that no baryon $\RAA$ exceeds this value. The uncertainties on the two terms are added in quadrature.
The resulting values of $\fprompt$ are about 0.93 and 0.81 for the $\Lc$ and $\Dzero$, respectively.

A summary of the systematic uncertainties on the corrected $\Lcplus$ and $\Dzero$ yields is shown in Tab.~\ref{Tab:SystUnc}. The $\Dzero$ systematic uncertainties on the particle identification (PID), tracking and cut variation are taken from Ref.~\cite{Acharya:2018hre} and are not discussed in the following.

\begin{table}[!t]
\begin{center}
\begin{tabular}{l c c}
Uncertainty & $\Lcplus$ & $\Dzero$\\
\hline
Raw-yield extraction & 8\% & 2\% \\
Tracking efficiency & 3.6\% & 5\%\\
PID & 5\% & negl.\\
Cut variation &2\% & 5\%\\
MC $\pt$ shape & 2\% & negl.\\
MC centrality weights & 3\% & negl.\\
Feed-down subtraction & $^{+6}_{-12}$\% &$^{+12}_{-13}$\%\\
Branching ratio & 5\% & 1\% \\
\end{tabular}
\end{center}
\caption{Systematic uncertainties on the corrected yields.  When the uncertainty was found to be $< 1$\%, it was considered negligible (negl��. in the table).}
\label{Tab:SystUnc}
\end{table}

The systematic uncertainty on the raw-yield extraction for $\Lc$ and $\Dzero$ was estimated by repeating the fits several times varying (i) the lower and upper limits of the fit range, (ii) the background fit function and (iii) only in the case of the $\Lc$, considering the Gaussian mean and width as free parameters in the fit. In addition, the signal yield was obtained by integrating the invariant-mass distribution after subtracting the background estimated from an exponential fit to the sidebands. 
 
For the $\Lc$, the systematic uncertainty on the tracking efficiency was evaluated by comparing the probability of matching tracks reconstructed in the TPC to ITS hits in data and simulation and by varying the quality cuts to select the tracks used in the analysis. The contribution due to the variation of the quality cuts was evaluated using protons from $\Lambda$ decays and an inclusive $\Kzeros$ sample and by calculating the ratio of the corrected yields obtained using different selection criteria. The uncertainty on the ITS-TPC matching efficiency is defined as the relative difference of the matching efficiency in data and simulations after weighting the relative abundances of primary and secondary particles in the simulations to match those in data. The latter were estimated via fits to the track impact-parameter distributions. The values calculated as a function of track momentum were propagated to the $\pt$-differential uncertainty of the $\Lc$ using a Monte Carlo simulation. A 3\% systematic uncertainty on the ITS-TPC matching efficiency of proton tracks was assigned while for the $\Kzeros$ the matching is not required. The uncertainty resulting from these studies was added in quadrature to the uncertainty on the track selection. 

The systematic uncertainty on the $\Lc$ PID efficiency was evaluated using protons from the decay of $\Lambda$ baryons. The ratio of the $\Lambda$ yield measured with PID to that measured without PID was calculated in both data and MC and their difference was used to estimate the systematic uncertainty.

Systematic uncertainties on the efficiencies can also arise from possible differences in the distributions and resolutions of selection variables between data and simulation. The systematic effect induced by these imperfections was estimated by repeating the analysis varying the main selection criteria for the candidates. 
The efficiencies determined from the simulations depend also on the generated $\pt$ distributions of the $\Lc$ and the $\Dzero$. The central values of the correction factors were obtained by re-weighting the $\Lc$ and $\Dzero$ distributions generated by PYTHIA as described above. For the $\Dzero$, the efficiencies calculated with and without the $\pt$ weights are compatible and therefore no uncertainty was assigned. 
For the $\Lc$, the systematic uncertainty was defined by considering the variation of the efficiencies determined with different generated $\pt$ shapes. The new $\Lc$ $\pt$ shape was calculated by multiplying the measured $\Dzero$ $\pt$ distribution with the $\LcOverDzero$ ratios predicted by the models~\cite{Oh:2009zj} and ~\cite{MartinezGarcia:2007hf}.

Finally, the efficiencies in the centrality class 0--80\% depend on the centrality weights used to combine the efficiencies in the smaller centrality classes. The stability of the efficiencies against the variation of the centrality weights was tested by recalculating the efficiencies without weighting for $\left< N_{\mathrm{coll}} \right>$ and, for the $\Lc$, using as an alternative centrality weight the product $\Lambda/ {\Kzeros} \cdot \left< N_{\mathrm{coll}} \right>$, where the ratio $\Lambda/ \Kzeros$ is taken from Ref.~\cite{Abelev:2013xaa}.

The systematic uncertainty on the subtraction of feed-down from beauty-hadron decays was estimated by varying (i) the $\pt$-differential cross section of feed-down $\Lc$ ($\Dzero$) from FONLL calculations within the theoretical uncertainties (see Ref.~\cite{Acharya:2017kfy} for details on the $\Lc$ and Ref.~\cite{ALICE:2011aa} for the $\Dzero$) and (ii) the ratio of prompt and feed-down $R_{\mathrm{AA}}$ as described above.

The production yields of $\Lc$ and $\Dzero$ also have a global systematic uncertainty due to the branching ratio. 

\section{Results}
\begin{figure}[!t]
  \begin{center}
    \includegraphics[angle=0, width=1.0\textwidth]{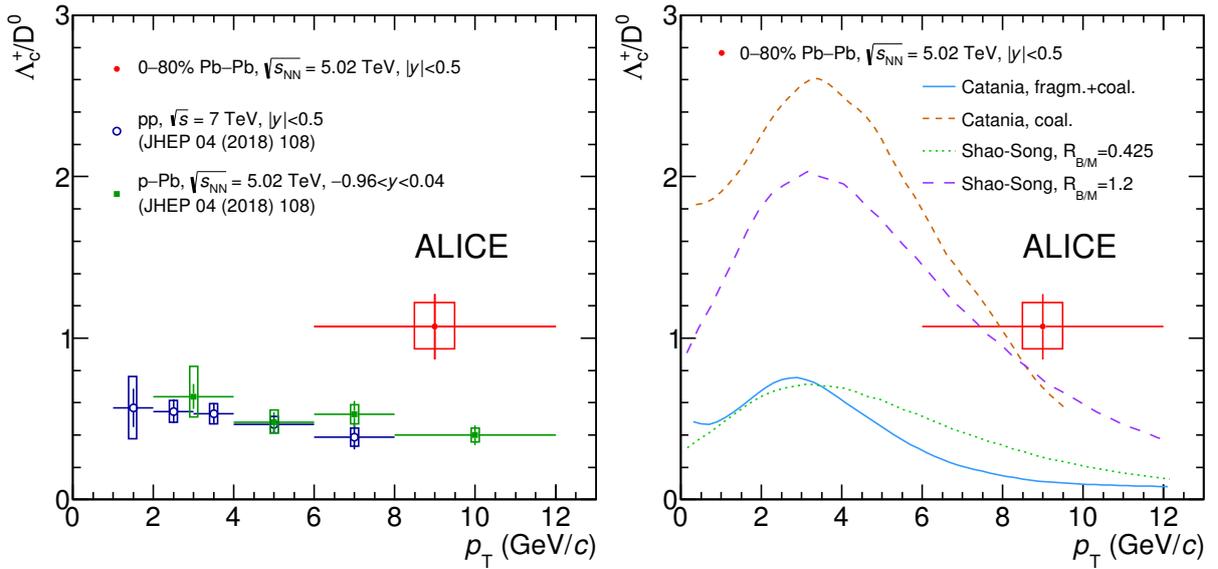}
  \end{center}
  \caption{$\LcOverDzero$ ratio as a function of $\pt$ in 0--80\% most central Pb--Pb collisions compared with the measurements in pp and p--Pb collisions~\cite{Acharya:2017kfy} (left), and model calculations~\cite{Plumari:2017ntm} (right). Statistical and systematic uncertainties are presented as vertical bars and boxes, respectively.}
  \label{fig:LcOverD0} 
\end{figure} 

The yield of prompt $\Lcplus$ baryons measured in Pb--Pb collisions at $\sqrtsNN = 5.02~\TeV$ in the 0--80\% centrality class in $|y|<0.5$ and $6<\pt < 12~\gevc$ 
is $N^{\Lcplus} = (2.1 \pm 0.4~{\rm(stat.)} ^{+0.3}_{-0.4}~{\rm (syst.)}) \times 10^{-2}$.

\begin{figure}[!h]
  \begin{center}
    \includegraphics[angle=0, width=1.0\textwidth]{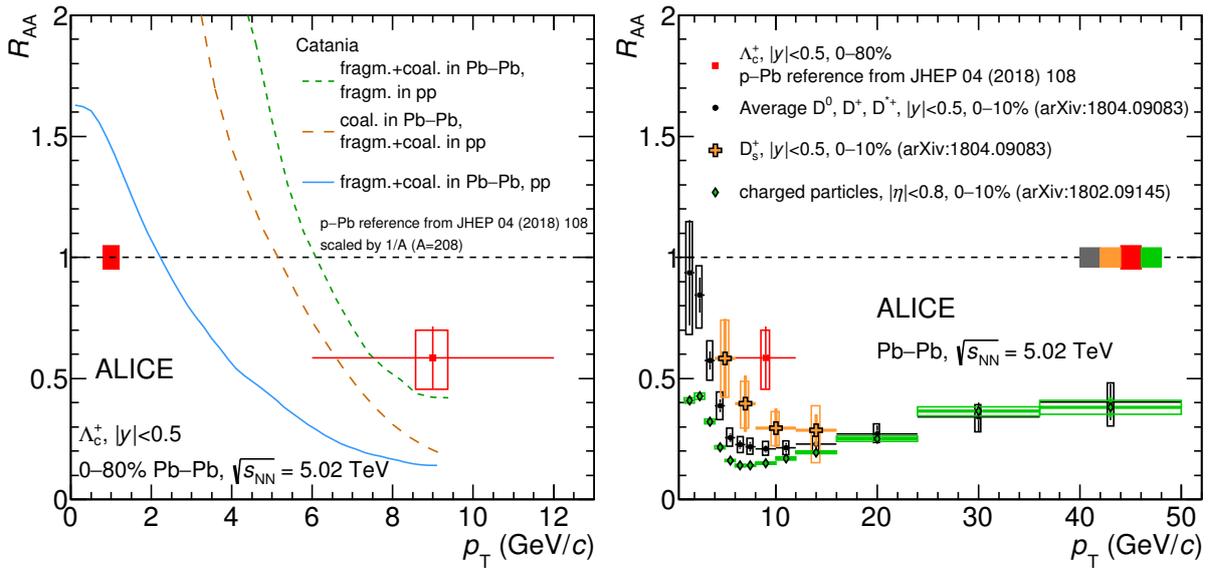}
  \end{center}
  \caption{$\RAA$ of prompt $\Lcplus$ compared with model calculations~\cite{Plumari:2017ntm,Li:2017zuj,Song:2018tpv} (left), and the non-strange D mesons, $\Ds$, and charged particle $\RAA$ in 0--10\% most central Pb--Pb collisions for $\pt > 1~\gevc$~\cite{Acharya:2018hre,Acharya:2018qsh} (right). Statistical, systematic and normalisation uncertainties are presented as vertical bars, empty boxes and shaded boxes around unity, respectively.}
  \label{fig:LcRAA} 
\end{figure} 

The measured $\LcOverDzero$ ratio is shown in Fig.~\ref{fig:LcOverD0}. The systematic uncertainty of the $\Lcplus$-baryon production arising from the tracking efficiency was treated as fully correlated to that of the $\Dzero$ meson. The contribution to the feed-down uncertainty related to heavy-quark energy loss and that originating from the FONLL uncertainty on the feed-down $\Lcplus$ and $\Dzero$ cross sections were treated as fully correlated when propagated to the ratio. All the other sources of uncertainty were considered as uncorrelated. In the left panel of Fig.~\ref{fig:LcOverD0}, the $\LcOverDzero$ ratio measured in $\PbPb$ collisions is compared with the results obtained by the ALICE Collaboration in minimum-bias pp and p--Pb collisions at $\sqrts= 7~\TeV$ and $\sqrtsNN=5.02~\TeV$~\cite{Acharya:2017kfy}, respectively.  The ratio measured in Pb--Pb collisions is higher than that measured in pp and p--Pb collisions. In particular, the values in p--Pb and Pb--Pb collisions differ by about two standard deviations of the combined statistical and systematic uncertainties in $6<\pt < 12~\gevc$. 

The $\LcOverDzero$ ratio in $\PbPb$ collisions is compared with theoretical model calculations in the right panel of Fig.~\ref{fig:LcOverD0}. The Catania model~\cite{Plumari:2017ntm} provides two different treatments of hadronisation. In one case, charm quarks hadronise via coalescence only. In the other case, a coalescence plus vacuum fragmentation modelling of hadronisation is considered: at increasing $\pt$ the coalescence probability decreases and eventually vacuum fragmentation takes over. For $\Dzero$ mesons, the shape of the fragmentation function is tuned assuring that the experimental results on D-meson production in pp collisions are well described by a fragmentation hadronisation mechanism. Data from $\mathrm {e^+e^-}$ collisions are used to fix the shape of the fragmentation functions for $\Lcplus$. The coalescence mechanism is treated as a three-quark process and implemented through the Wigner formalism. The momentum spectrum of hadrons formed by coalescence is obtained from the quark phase-space distributions and the hadron wave function. The width parameters of the hadron wave functions are calculated from the charge radius of the hadrons according to the quark model. The hadron wave function normalisation is determined by requiring a total coalescence probability for charm quarks equal to unity for zero-momentum heavy quarks. Moreover, the contributions from the first excited states for D and $\Lc$ hadrons were included in the calculations. The experimental results are described by the model calculation including coalescence only. The curve obtained by modelling charm hadronisation via vacuum fragmentation plus coalescence, which describes the $\LcOverDzero$ ratio measured in \AuAu collisions at RHIC energy~\cite{Xie:2017jcq}, significantly underestimates the measurement in \PbPb collisions at the LHC.
In the Shao-Song model~\cite{Li:2017zuj,Song:2018tpv}, coalescence involves quarks which are close in momentum space, and it takes place mainly for the quark with a given fraction of the momentum of the hadron. It does not consider the Wigner formalism to describe the spatial and momentum distribution of quarks in a hadron. It can not directly predict the absolute magnitude of the $\LcOverDzero$ ratio because the relative production of single-charm baryons and single-charm mesons ${\rm R_{BM}}$ is treated as a parameter of the model. The curve obtained by considering ${\rm R_{BM}}=0.425$, which is the value needed to describe the results in pp and p--Pb collisions, underestimates the $\LcOverDzero$ ratio measured in Pb--Pb collisions. An ${\rm R_{BM}}=1.2$ is needed to achieve a better description of the experimental results in \PbPb collisions. However, the hadronisation mechanism via quark coalescence included in the model is responsible of the $\pt$ dependence of the $\LcOverDzero$ ratio, which needs to be verified by comparing to a measurement at lower $\pt$. 
The $\RAA$ of prompt $\Lcplus$ was obtained by considering as reference the $\Lcplus$ cross section measured in p--Pb collisions at $\sqrtsNN = 5.02~\TeV$~\cite{Acharya:2017kfy} scaled by $1/A$ ($A=208$) and corrected for the different rapidity coverage of the p--Pb measurement. The cross section measured in p--Pb was scaled in each $\pt$ interval to $|y|<0.5$ using a correction factor obtained with FONLL  calculations~\cite{Cacciari:1998it,Cacciari:2001td}. The correction factor was determined from the ratios of the cross sections calculated with FONLL in the rapidity intervals $|y|<0.5$ and $-0.96<y<0.04$. Since FONLL does not provide predictions for $\Lcplus$ baryons, the average of the correction factors obtained for $\Dzero$, ${\rm D}^{+}$ and bare charm quarks, which was found to be $1.024 \pm 0.008$, was used. The choice of using the p--Pb cross section to obtain the reference for the $\RAA$ was motivated by the fact that it was measured up to $\pt=12~\gevc$, while the measurement in pp collisions at $\sqrts = 7~\TeV$ in $|y|<0.5$ only reaches $\pt=8~\gevc$. In addition, the $\Lcplus$ nuclear modification factor measured in p--Pb collisions is consistent with unity for $\pt>2~\gevc$~\cite{Acharya:2017kfy}. 
The $\Lcplus$ reference cross section in $6<\pt < 12~\gevc$ was obtained by combining the results in the transverse momentum intervals $6<\pt <8~\gevc$ and $8<\pt < 12~\gevc$. The uncertainties were propagated treating the statistical and the systematic uncertainties on the yield extraction as uncorrelated and the other sources of systematic uncertainty as correlated in $\pt$. The $\Lcplus$ $\Raa$ also has a 3.75\% uncertainty due to the normalisation of the $\Lcplus$ p--Pb cross section at $\sqrtsNN = 5.02$~TeV~\cite{Acharya:2017kfy} and a 2.4\% uncertainty on the average nuclear overlap function $\left< \TAA \right>$, which were added in quadrature.
In the left panel of Fig.~\ref{fig:LcRAA}, the $\RAA$ of prompt $\Lcplus$ is compared with Catania model calculations~\cite{Plumari:2017ntm}.
The three curves are obtained by considering different treatments of the hadronisation mechanisms in pp and Pb--Pb collisions. The short-dashed curve represents the $\Lcplus$ $\RAA$ as obtained by including both vacuum fragmentation and quark coalescence for charm hadronisation in Pb--Pb and only fragmentation in pp collisions. The long-dashed curve includes only coalescence in Pb--Pb and fragmentation plus coalescence in pp collisions. The solid curve is obtained by considering fragmentation plus coalescence in both collision systems. The limited precision and the large $\pt$ interval of this first measurement prevent us to draw a firm conclusion on which combination of the hadronisation mechanisms in the two collision systems better describes the result. Moreover, the comparison between the different scenarios obtained from the Catania model demonstrates that it is crucial to also understand the $\Lcplus$ production mechanism in pp collisions to interpret the $\RAA$ measurement.
The right panel of Fig.~\ref{fig:LcRAA} shows the $\RAA$ of prompt $\Lcplus$ baryons measured in the 0--80\% centrality class (that is dominated by the 0--10\% production given the scaling of the yields with $N_{\rm coll}\cdot R_{\rm AA}$) compared with the average nuclear modification factors of non-strange D mesons,  $\Ds$ mesons, and charged particles measured in the 0--10\% centrality class~\cite{Acharya:2018hre}. The $\RAA$ of charged particles is smaller than that of D mesons by more than $2\sigma$ of the combined statistical and systematic uncertainties up to $\pt=8~\gevc$, while they are compatible within $1\sigma$ for $\pt>10~\gevc$. The $\RAA$ values of $\Ds$ mesons are larger than those of non-strange D mesons, but the two measurements are compatible within one standard deviation of the combined uncertainties~\cite{Acharya:2018hre}. A hint of a larger $\Lcplus$ $\RAA$ with respect to non-strange D mesons is observed, although the results are compared for different centrality classes. 
A $\Dzero$ $\RAA= 0.27 \pm 0.01 {\rm (stat.) \pm 0.04 {\rm (syst.)}}$ was measured in $6<\pt<12~\gevc$ in the 0--80\% centrality class. The $\Dzero$ $\RAA$ has also a 3.5\% uncertainty arising from the normalisation of the cross section measured in pp collisions at $\sqrts=7$~TeV, and a 2.4\% uncertainty on the average nuclear overlap function $\left< \TAA \right>$. The $\pt$-differential cross section of prompt $\Dzero$ mesons with $|y|<0.5$ in pp collisions at $\sqrts=5.02$ TeV, used as reference for the nuclear modification factor, was obtained by scaling the measurement at $\sqrts=7$ TeV~\cite{Acharya:2017jgo} to $\sqrts=5.02$ TeV using FONLL calculations~\cite{Cacciari:1998it,Cacciari:2001td}. The scaling was applied to the $\Dzero$ cross section obtained in $6<\pt<12~\gevc$ by combining the results in the $\pt$ intervals of the measurement at $\sqrts=7$~TeV. The statistical and the systematic uncertainties on the yield extraction were propagated as uncorrelated. The other contributions to the systematic uncertainty were considered as fully correlated among the $\pt$ intervals.
A difference of about 1.7$\sigma$ is obtained when comparing the $\Lcplus$ $\RAA$ with that of the $\Dzero$ in $6<\pt<12~\gevc$ and 0--80\% centrality interval. This observation is qualitatively in agreement with a scenario where a significant fraction of charm quarks hadronise via coalescence with light quarks from the medium leading to an enhanced baryon production with respect to that of mesons.  

 \section{Summary}
The measurement of the production of prompt $\Lcplus$ baryons in the 0--80\% most central Pb--Pb collisions at $\sqrtsNN = 5.02~\TeV$ was presented. The result was obtained at midrapidity, $|y|<0.5$, in the $6<\pt<12~\gevc$ transverse momentum interval. The $\LcOverDzero$ ratio is larger than the ratio measured in pp and p--Pb collisions at $\sqrts = 7~\TeV$ and $\sqrtsNN=5.02~\TeV$~\cite{Acharya:2017kfy}, respectively. The $\LcOverDzero$ ratio measured in Pb--Pb collisions is described by a model calculation implementing only charm quark hadronisation via quark coalescence and it is underestimated when also vacuum fragmentation is included. The comparison of the $\Lcplus$ nuclear modification factor with non-strange D and $\Ds$ meson results, which were measured in 0--10\% most central Pb--Pb collisions, suggests a hint of a hierarchy, conceivable in a scenario where charm quark hadronisation can occur via coalescence processes, thus enhancing the $\Lcplus$-baryon and $\Ds$-meson production with respect to non-strange D mesons. However, the limited precision of this first measurement prevents us from drawing a firm conclusion.

A higher precision for a $\Lcplus$-baryon production measurement with finer granularity in $\pt$ and centrality will be achieved with future datasets to be collected during LHC Run~2 and, in particular, during the LHC Run~3 and 4, following the major upgrade of the ALICE apparatus~\cite{AliceUpgrade,Abelev:2014dna}.

\newenvironment{acknowledgement}{\relax}{\relax}
\begin{acknowledgement}
\section*{Acknowledgements}
\input{fa_2018-09-19.tex}
\end{acknowledgement}

\bibliographystyle{utphys}   \bibliography{LcinPbPb_paper}

\newpage
\appendix
\section{The ALICE Collaboration}
\label{app:collab}
\input{Alice_Authorlist_2018-Sep-19.tex}
\end{document}

%% file: fa_2018-09-19.tex

The ALICE Collaboration would like to thank all its engineers and technicians for their invaluable contributions to the construction of the experiment and the CERN accelerator teams for the outstanding performance of the LHC complex.
The ALICE Collaboration gratefully acknowledges the resources and support provided by all Grid centres and the Worldwide LHC Computing Grid (WLCG) collaboration.
The ALICE Collaboration acknowledges the following funding agencies for their support in building and running the ALICE detector:
A. I. Alikhanyan National Science Laboratory (Yerevan Physics Institute) Foundation (ANSL), State Committee of Science and World Federation of Scientists (WFS), Armenia;
Austrian Academy of Sciences and Nationalstiftung f\"{u}r Forschung, Technologie und Entwicklung, Austria;
Ministry of Communications and High Technologies, National Nuclear Research Center, Azerbaijan;
Conselho Nacional de Desenvolvimento Cient\'{\i}fico e Tecnol\'{o}gico (CNPq), Universidade Federal do Rio Grande do Sul (UFRGS), Financiadora de Estudos e Projetos (Finep) and Funda\c{c}\~{a}o de Amparo \`{a} Pesquisa do Estado de S\~{a}o Paulo (FAPESP), Brazil;
Ministry of Science \& Technology of China (MSTC), National Natural Science Foundation of China (NSFC) and Ministry of Education of China (MOEC) , China;
Ministry of Science and Education, Croatia;
Centro de Aplicaciones Tecnol\'{o}gicas y Desarrollo Nuclear (CEADEN), Cubaenerg\'{\i}a, Cuba;
Ministry of Education, Youth and Sports of the Czech Republic, Czech Republic;
The Danish Council for Independent Research | Natural Sciences, the Carlsberg Foundation and Danish National Research Foundation (DNRF), Denmark;
Helsinki Institute of Physics (HIP), Finland;
Commissariat \`{a} l'Energie Atomique (CEA) and Institut National de Physique Nucl\'{e}aire et de Physique des Particules (IN2P3) and Centre National de la Recherche Scientifique (CNRS), France;
Bundesministerium f\"{u}r Bildung, Wissenschaft, Forschung und Technologie (BMBF) and GSI Helmholtzzentrum f\"{u}r Schwerionenforschung GmbH, Germany;
General Secretariat for Research and Technology, Ministry of Education, Research and Religions, Greece;
National Research, Development and Innovation Office, Hungary;
Department of Atomic Energy Government of India (DAE), Department of Science and Technology, Government of India (DST), University Grants Commission, Government of India (UGC) and Council of Scientific and Industrial Research (CSIR), India;
Indonesian Institute of Science, Indonesia;
Centro Fermi - Museo Storico della Fisica e Centro Studi e Ricerche Enrico Fermi and Istituto Nazionale di Fisica Nucleare (INFN), Italy;
Institute for Innovative Science and Technology , Nagasaki Institute of Applied Science (IIST), Japan Society for the Promotion of Science (JSPS) KAKENHI and Japanese Ministry of Education, Culture, Sports, Science and Technology (MEXT), Japan;
Consejo Nacional de Ciencia (CONACYT) y Tecnolog\'{i}a, through Fondo de Cooperaci\'{o}n Internacional en Ciencia y Tecnolog\'{i}a (FONCICYT) and Direcci\'{o}n General de Asuntos del Personal Academico (DGAPA), Mexico;
Nederlandse Organisatie voor Wetenschappelijk Onderzoek (NWO), Netherlands;
The Research Council of Norway, Norway;
Commission on Science and Technology for Sustainable Development in the South (COMSATS), Pakistan;
Pontificia Universidad Cat\'{o}lica del Per\'{u}, Peru;
Ministry of Science and Higher Education and National Science Centre, Poland;
Korea Institute of Science and Technology Information and National Research Foundation of Korea (NRF), Republic of Korea;
Ministry of Education and Scientific Research, Institute of Atomic Physics and Romanian National Agency for Science, Technology and Innovation, Romania;
Joint Institute for Nuclear Research (JINR), Ministry of Education and Science of the Russian Federation and National Research Centre Kurchatov Institute, Russia;
Ministry of Education, Science, Research and Sport of the Slovak Republic, Slovakia;
National Research Foundation of South Africa, South Africa;
Swedish Research Council (VR) and Knut \& Alice Wallenberg Foundation (KAW), Sweden;
European Organization for Nuclear Research, Switzerland;
National Science and Technology Development Agency (NSDTA), Suranaree University of Technology (SUT) and Office of the Higher Education Commission under NRU project of Thailand, Thailand;
Turkish Atomic Energy Agency (TAEK), Turkey;
National Academy of  Sciences of Ukraine, Ukraine;
Science and Technology Facilities Council (STFC), United Kingdom;
National Science Foundation of the United States of America (NSF) and United States Department of Energy, Office of Nuclear Physics (DOE NP), United States of America.

%% file: Alice_Authorlist_2018-Sep-19.tex

\begingroup
\small
\begin{flushleft}
S.~Acharya\Irefn{org140}\And 
F.T.-.~Acosta\Irefn{org20}\And 
D.~Adamov\'{a}\Irefn{org93}\And 
S.P.~Adhya\Irefn{org140}\And 
A.~Adler\Irefn{org74}\And 
J.~Adolfsson\Irefn{org80}\And 
M.M.~Aggarwal\Irefn{org98}\And 
G.~Aglieri Rinella\Irefn{org34}\And 
M.~Agnello\Irefn{org31}\And 
N.~Agrawal\Irefn{org48}\And 
Z.~Ahammed\Irefn{org140}
% ...
\And
\renewcommand\labelenumi{\textsuperscript{\theenumi}~}

\section*{Affiliation notes}
\renewcommand\theenumi{\roman{enumi}}
\begin{Authlist}
\item \Adef{org*}Deceased
\item \Adef{orgI}Dipartimento DET del Politecnico di Torino, Turin, Italy
\item \Adef{orgII}M.V. Lomonosov Moscow State University, D.V. Skobeltsyn Institute of Nuclear, Physics, Moscow, Russia
\item \Adef{orgIII}Department of Applied Physics, Aligarh Muslim University, Aligarh, India
\item \Adef{orgIV}Institute of Theoretical Physics, University of Wroclaw, Poland
\end{Authlist}

\section*{Collaboration Institutes}
\renewcommand\theenumi{\arabic{enumi}~}
\begin{Authlist}
\item \Idef{org1}A.I. Alikhanyan National Science Laboratory (Yerevan Physics Institute) Foundation, Yerevan, Armenia
\item \Idef{org2}Bogolyubov Institute for Theoretical Physics, National Academy of Sciences of Ukraine, Kiev, Ukraine
\item \Idef{org3}Bose Institute, Department of Physics  and Centre for Astroparticle Physics and Space Science (CAPSS), Kolkata, India
\item \Idef{org4}Budker Institute for Nuclear Physics, Novosibirsk, Russia
\item \Idef{org5}California Polytechnic State University, San Luis Obispo, California, United States
\item \Idef{org6}Central China Normal University, Wuhan, China
\item \Idef{org7}Centre de Calcul de l'IN2P3, Villeurbanne, Lyon, France
\item \Idef{org8}Centro de Aplicaciones Tecnol\'{o}gicas y Desarrollo Nuclear (CEADEN), Havana, Cuba
\item \Idef{org9}Centro de Investigaci\'{o}n y de Estudios Avanzados (CINVESTAV), Mexico City and M\'{e}rida, Mexico
\item \Idef{org10}Centro Fermi - Museo Storico della Fisica e Centro Studi e Ricerche ``Enrico Fermi', Rome, Italy
\item \Idef{org11}Chicago State University, Chicago, Illinois, United States
\item \Idef{org12}China Institute of Atomic Energy, Beijing, China
\item \Idef{org13}Chonbuk National University, Jeonju, Republic of Korea
\item \Idef{org14}Comenius University Bratislava, Faculty of Mathematics, Physics and Informatics, Bratislava, Slovakia
\item \Idef{org15}COMSATS Institute of Information Technology (CIIT), Islamabad, Pakistan
\item \Idef{org16}Creighton University, Omaha, Nebraska, United States
\item \Idef{org17}Department of Physics, Aligarh Muslim University, Aligarh, India
\item \Idef{org18}Department of Physics, Pusan National University, Pusan, Republic of Korea
\item \Idef{org19}Department of Physics, Sejong University, Seoul, Republic of Korea
\item \Idef{org20}Department of Physics, University of California, Berkeley, California, United States
\item \Idef{org21}Department of Physics, University of Oslo, Oslo, Norway
\item \Idef{org22}Department of Physics and Technology, University of Bergen, Bergen, Norway
\item \Idef{org23}Dipartimento di Fisica dell'Universit\`{a} 'La Sapienza' and Sezione INFN, Rome, Italy
\item \Idef{org24}Dipartimento di Fisica dell'Universit\`{a} and Sezione INFN, Cagliari, Italy
\item \Idef{org25}Dipartimento di Fisica dell'Universit\`{a} and Sezione INFN, Trieste, Italy
\item \Idef{org26}Dipartimento di Fisica dell'Universit\`{a} and Sezione INFN, Turin, Italy
\item \Idef{org27}Dipartimento di Fisica e Astronomia dell'Universit\`{a} and Sezione INFN, Bologna, Italy
\item \Idef{org28}Dipartimento di Fisica e Astronomia dell'Universit\`{a} and Sezione INFN, Catania, Italy
\item \Idef{org29}Dipartimento di Fisica e Astronomia dell'Universit\`{a} and Sezione INFN, Padova, Italy
\item \Idef{org30}Dipartimento di Fisica `E.R.~Caianiello' dell'Universit\`{a} and Gruppo Collegato INFN, Salerno, Italy
\item \Idef{org31}Dipartimento DISAT del Politecnico and Sezione INFN, Turin, Italy
\item \Idef{org32}Dipartimento di Scienze e Innovazione Tecnologica dell'Universit\`{a} del Piemonte Orientale and INFN Sezione di Torino, Alessandria, Italy
\item \Idef{org33}Dipartimento Interateneo di Fisica `M.~Merlin' and Sezione INFN, Bari, Italy
\item \Idef{org34}European Organization for Nuclear Research (CERN), Geneva, Switzerland
\item \Idef{org35}Faculty of Electrical Engineering, Mechanical Engineering and Naval Architecture, University of Split, Split, Croatia
\item \Idef{org36}Faculty of Engineering and Science, Western Norway University of Applied Sciences, Bergen, Norway
\item \Idef{org37}Faculty of Nuclear Sciences and Physical Engineering, Czech Technical University in Prague, Prague, Czech Republic
\item \Idef{org38}Faculty of Science, P.J.~\v{S}af\'{a}rik University, Ko\v{s}ice, Slovakia
\item \Idef{org39}Frankfurt Institute for Advanced Studies, Johann Wolfgang Goethe-Universit\"{a}t Frankfurt, Frankfurt, Germany
\item \Idef{org40}Gangneung-Wonju National University, Gangneung, Republic of Korea
\item \Idef{org41}Gauhati University, Department of Physics, Guwahati, India
\item \Idef{org42}Helmholtz-Institut f\"{u}r Strahlen- und Kernphysik, Rheinische Friedrich-Wilhelms-Universit\"{a}t Bonn, Bonn, Germany
\item \Idef{org43}Helsinki Institute of Physics (HIP), Helsinki, Finland
\item \Idef{org44}High Energy Physics Group,  Universidad Aut\'{o}noma de Puebla, Puebla, Mexico
\item \Idef{org45}Hiroshima University, Hiroshima, Japan
\item \Idef{org46}Hochschule Worms, Zentrum  f\"{u}r Technologietransfer und Telekommunikation (ZTT), Worms, Germany
\item \Idef{org47}Horia Hulubei National Institute of Physics and Nuclear Engineering, Bucharest, Romania
\item \Idef{org48}Indian Institute of Technology Bombay (IIT), Mumbai, India
\item \Idef{org49}Indian Institute of Technology Indore, Indore, India
\item \Idef{org50}Indonesian Institute of Sciences, Jakarta, Indonesia
\item \Idef{org51}INFN, Laboratori Nazionali di Frascati, Frascati, Italy
\item \Idef{org52}INFN, Sezione di Bari, Bari, Italy
\item \Idef{org53}INFN, Sezione di Bologna, Bologna, Italy
\item \Idef{org54}INFN, Sezione di Cagliari, Cagliari, Italy
\item \Idef{org55}INFN, Sezione di Catania, Catania, Italy
\item \Idef{org56}INFN, Sezione di Padova, Padova, Italy
\item \Idef{org57}INFN, Sezione di Roma, Rome, Italy
\item \Idef{org58}INFN, Sezione di Torino, Turin, Italy
\item \Idef{org59}INFN, Sezione di Trieste, Trieste, Italy
\item \Idef{org60}Inha University, Incheon, Republic of Korea
\item \Idef{org61}Institut de Physique Nucl\'{e}aire d'Orsay (IPNO), Institut National de Physique Nucl\'{e}aire et de Physique des Particules (IN2P3/CNRS), Universit\'{e} de Paris-Sud, Universit\'{e} Paris-Saclay, Orsay, France
\item \Idef{org62}Institute for Nuclear Research, Academy of Sciences, Moscow, Russia
\item \Idef{org63}Institute for Subatomic Physics, Utrecht University/Nikhef, Utrecht, Netherlands
\item \Idef{org64}Institute for Theoretical and Experimental Physics, Moscow, Russia
\item \Idef{org65}Institute of Experimental Physics, Slovak Academy of Sciences, Ko\v{s}ice, Slovakia
\item \Idef{org66}Institute of Physics, Homi Bhabha National Institute, Bhubaneswar, India
\item \Idef{org67}Institute of Physics of the Czech Academy of Sciences, Prague, Czech Republic
\item \Idef{org68}Institute of Space Science (ISS), Bucharest, Romania
\item \Idef{org69}Institut f\"{u}r Kernphysik, Johann Wolfgang Goethe-Universit\"{a}t Frankfurt, Frankfurt, Germany
\item \Idef{org70}Instituto de Ciencias Nucleares, Universidad Nacional Aut\'{o}noma de M\'{e}xico, Mexico City, Mexico
\item \Idef{org71}Instituto de F\'{i}sica, Universidade Federal do Rio Grande do Sul (UFRGS), Porto Alegre, Brazil
\item \Idef{org72}Instituto de F\'{\i}sica, Universidad Nacional Aut\'{o}noma de M\'{e}xico, Mexico City, Mexico
\item \Idef{org73}iThemba LABS, National Research Foundation, Somerset West, South Africa
\item \Idef{org74}Johann-Wolfgang-Goethe Universit\"{a}t Frankfurt Institut f\"{u}r Informatik, Fachbereich Informatik und Mathematik, Frankfurt, Germany
\item \Idef{org75}Joint Institute for Nuclear Research (JINR), Dubna, Russia
\item \Idef{org76}Korea Institute of Science and Technology Information, Daejeon, Republic of Korea
\item \Idef{org77}KTO Karatay University, Konya, Turkey
\item \Idef{org78}Laboratoire de Physique Subatomique et de Cosmologie, Universit\'{e} Grenoble-Alpes, CNRS-IN2P3, Grenoble, France
\item \Idef{org79}Lawrence Berkeley National Laboratory, Berkeley, California, United States
\item \Idef{org80}Lund University Department of Physics, Division of Particle Physics, Lund, Sweden
\item \Idef{org81}Nagasaki Institute of Applied Science, Nagasaki, Japan
\item \Idef{org82}Nara Women{'}s University (NWU), Nara, Japan
\item \Idef{org83}National and Kapodistrian University of Athens, School of Science, Department of Physics , Athens, Greece
\item \Idef{org84}National Centre for Nuclear Research, Warsaw, Poland
\item \Idef{org85}National Institute of Science Education and Research, Homi Bhabha National Institute, Jatni, India
\item \Idef{org86}National Nuclear Research Center, Baku, Azerbaijan
\item \Idef{org87}National Research Centre Kurchatov Institute, Moscow, Russia
\item \Idef{org88}Niels Bohr Institute, University of Copenhagen, Copenhagen, Denmark
\item \Idef{org89}Nikhef, National institute for subatomic physics, Amsterdam, Netherlands
\item \Idef{org90}NRC Kurchatov Institute IHEP, Protvino, Russia
\item \Idef{org91}NRNU Moscow Engineering Physics Institute, Moscow, Russia
\item \Idef{org92}Nuclear Physics Group, STFC Daresbury Laboratory, Daresbury, United Kingdom
\item \Idef{org93}Nuclear Physics Institute of the Czech Academy of Sciences, \v{R}e\v{z} u Prahy, Czech Republic
\item \Idef{org94}Oak Ridge National Laboratory, Oak Ridge, Tennessee, United States
\item \Idef{org95}Ohio State University, Columbus, Ohio, United States
\item \Idef{org96}Petersburg Nuclear Physics Institute, Gatchina, Russia
\item \Idef{org97}Physics department, Faculty of science, University of Zagreb, Zagreb, Croatia
\item \Idef{org98}Physics Department, Panjab University, Chandigarh, India
\item \Idef{org99}Physics Department, University of Jammu, Jammu, India
\item \Idef{org100}Physics Department, University of Rajasthan, Jaipur, India
\item \Idef{org101}Physikalisches Institut, Eberhard-Karls-Universit\"{a}t T\"{u}bingen, T\"{u}bingen, Germany
\item \Idef{org102}Physikalisches Institut, Ruprecht-Karls-Universit\"{a}t Heidelberg, Heidelberg, Germany
\item \Idef{org103}Physik Department, Technische Universit\"{a}t M\"{u}nchen, Munich, Germany
\item \Idef{org104}Research Division and ExtreMe Matter Institute EMMI, GSI Helmholtzzentrum f\"ur Schwerionenforschung GmbH, Darmstadt, Germany
\item \Idef{org105}Rudjer Bo\v{s}kovi\'{c} Institute, Zagreb, Croatia
\item \Idef{org106}Russian Federal Nuclear Center (VNIIEF), Sarov, Russia
\item \Idef{org107}Saha Institute of Nuclear Physics, Homi Bhabha National Institute, Kolkata, India
\item \Idef{org108}School of Physics and Astronomy, University of Birmingham, Birmingham, United Kingdom
\item \Idef{org109}Secci\'{o}n F\'{\i}sica, Departamento de Ciencias, Pontificia Universidad Cat\'{o}lica del Per\'{u}, Lima, Peru
\item \Idef{org110}Shanghai Institute of Applied Physics, Shanghai, China
\item \Idef{org111}St. Petersburg State University, St. Petersburg, Russia
\item \Idef{org112}Stefan Meyer Institut f\"{u}r Subatomare Physik (SMI), Vienna, Austria
\item \Idef{org113}SUBATECH, IMT Atlantique, Universit\'{e} de Nantes, CNRS-IN2P3, Nantes, France
\item \Idef{org114}Suranaree University of Technology, Nakhon Ratchasima, Thailand
\item \Idef{org115}Technical University of Ko\v{s}ice, Ko\v{s}ice, Slovakia
\item \Idef{org116}Technische Universit\"{a}t M\"{u}nchen, Excellence Cluster 'Universe', Munich, Germany
\item \Idef{org117}The Henryk Niewodniczanski Institute of Nuclear Physics, Polish Academy of Sciences, Cracow, Poland
\item \Idef{org118}The University of Texas at Austin, Austin, Texas, United States
\item \Idef{org119}Universidad Aut\'{o}noma de Sinaloa, Culiac\'{a}n, Mexico
\item \Idef{org120}Universidade de S\~{a}o Paulo (USP), S\~{a}o Paulo, Brazil
\item \Idef{org121}Universidade Estadual de Campinas (UNICAMP), Campinas, Brazil
\item \Idef{org122}Universidade Federal do ABC, Santo Andre, Brazil
\item \Idef{org123}University College of Southeast Norway, Tonsberg, Norway
\item \Idef{org124}University of Cape Town, Cape Town, South Africa
\item \Idef{org125}University of Houston, Houston, Texas, United States
\item \Idef{org126}University of Jyv\"{a}skyl\"{a}, Jyv\"{a}skyl\"{a}, Finland
\item \Idef{org127}University of Liverpool, Liverpool, United Kingdom
\item \Idef{org128}University of Science and Techonology of China, Hefei, China
\item \Idef{org129}University of Tennessee, Knoxville, Tennessee, United States
\item \Idef{org130}University of the Witwatersrand, Johannesburg, South Africa
\item \Idef{org131}University of Tokyo, Tokyo, Japan
\item \Idef{org132}University of Tsukuba, Tsukuba, Japan
\item \Idef{org133}Universit\'{e} Clermont Auvergne, CNRS/IN2P3, LPC, Clermont-Ferrand, France
\item \Idef{org134}Universit\'{e} de Lyon, Universit\'{e} Lyon 1, CNRS/IN2P3, IPN-Lyon, Villeurbanne, Lyon, France
\item \Idef{org135}Universit\'{e} de Strasbourg, CNRS, IPHC UMR 7178, F-67000 Strasbourg, France, Strasbourg, France
\item \Idef{org136} Universit\'{e} Paris-Saclay Centre d¿\'Etudes de Saclay (CEA), IRFU, Department de Physique Nucl\'{e}aire (DPhN), Saclay, France
\item \Idef{org137}Universit\`{a} degli Studi di Foggia, Foggia, Italy
\item \Idef{org138}Universit\`{a} degli Studi di Pavia, Pavia, Italy
\item \Idef{org139}Universit\`{a} di Brescia, Brescia, Italy
\item \Idef{org140}Variable Energy Cyclotron Centre, Homi Bhabha National Institute, Kolkata, India
\item \Idef{org141}Warsaw University of Technology, Warsaw, Poland
\item \Idef{org142}Wayne State University, Detroit, Michigan, United States
\item \Idef{org143}Westf\"{a}lische Wilhelms-Universit\"{a}t M\"{u}nster, Institut f\"{u}r Kernphysik, M\"{u}nster, Germany
\item \Idef{org144}Wigner Research Centre for Physics, Hungarian Academy of Sciences, Budapest, Hungary
\item \Idef{org145}Yale University, New Haven, Connecticut, United States
\item \Idef{org146}Yonsei University, Seoul, Republic of Korea
\end{Authlist}
\endgroup

%% file: LcinPbPb_paper_final.bbl
\providecommand{\href}[2]{#2}\begingroup\raggedright\begin{thebibliography}{10}

\bibitem{Braun-Munzinger:2015hba}
P.~Braun-Munzinger, V.~Koch, T.~Schaefer, and J.~Stachel, ``{Properties of hot
  and dense matter from relativistic heavy ion collisions}'',
  \href{http://dx.doi.org/10.1016/j.physrep.2015.12.003}{{\em Phys. Rept.}
  {\bfseries 621} (2016) 76--126},
\href{http://arxiv.org/abs/1510.00442}{{\ttfamily arXiv:1510.00442 [nucl-th]}}.

\bibitem{PhysRevD.90.094503}
A.~Bazavov {\em et~al.}, ``Equation of state in ($2+1$)-flavor $\rm{QCD}$'',
  \href{http://dx.doi.org/10.1103/PhysRevD.90.094503}{{\em Phys. Rev. D}
  {\bfseries 90} (2014) 094503}.
  \url{https://link.aps.org/doi/10.1103/PhysRevD.90.094503}.

\bibitem{Andronic:2015wma}
A.~Andronic {\em et~al.}, ``{Heavy-flavour and quarkonium production in the LHC
  era: from proton–proton to heavy-ion collisions}'',
  \href{http://dx.doi.org/10.1140/epjc/s10052-015-3819-5}{{\em Eur. Phys. J.}
  {\bfseries C76} no.~3, (2016) 107},
\href{http://arxiv.org/abs/1506.03981}{{\ttfamily arXiv:1506.03981 [nucl-ex]}}.

\bibitem{Prino:2016cni}
F.~Prino and R.~Rapp, ``{Open Heavy Flavor in QCD Matter and in Nuclear
  Collisions}'', \href{http://dx.doi.org/10.1088/0954-3899/43/9/093002}{{\em J.
  Phys.} {\bfseries G43} no.~9, (2016) 093002},
\href{http://arxiv.org/abs/1603.00529}{{\ttfamily arXiv:1603.00529 [nucl-ex]}}.

\bibitem{Greco:2003vf}
V.~Greco, C.~M. Ko, and R.~Rapp, ``{Quark coalescence for charmed mesons in
  ultrarelativistic heavy ion collisions}'',
  \href{http://dx.doi.org/10.1016/j.physletb.2004.06.064}{{\em Phys. Lett.}
  {\bfseries B595} (2004) 202--208},
\href{http://arxiv.org/abs/nucl-th/0312100}{{\ttfamily arXiv:nucl-th/0312100
  [nucl-th]}}.

\bibitem{Oh:2009zj}
Y.~Oh, C.~M. Ko, S.~H. Lee, and S.~Yasui, ``{Heavy baryon/meson ratios in
  relativistic heavy ion collisions}'',
  \href{http://dx.doi.org/10.1103/PhysRevC.79.044905}{{\em Phys. Rev.}
  {\bfseries C79} (2009) 044905},
\href{http://arxiv.org/abs/0901.1382}{{\ttfamily arXiv:0901.1382 [nucl-th]}}.

\bibitem{Plumari:2017ntm}
S.~Plumari, V.~Minissale, S.~K. Das, G.~Coci, and V.~Greco, ``{Charmed Hadrons
  from Coalescence plus Fragmentation in relativistic nucleus-nucleus
  collisions at RHIC and LHC}'',
  \href{http://dx.doi.org/10.1140/epjc/s10052-018-5828-7}{{\em Eur. Phys. J.}
  {\bfseries C78} no.~4, (2018) 348},
\href{http://arxiv.org/abs/1712.00730}{{\ttfamily arXiv:1712.00730 [hep-ph]}}.

\bibitem{Abelev:2013xaa}
{\bfseries ALICE} Collaboration, B.~Abelev {\em et~al.}, ``{${\rm K^0_S}$ and
  $\Lambda$ production in Pb-Pb collisions at $\sqrt{s_{NN}}$ = 2.76 TeV}'',
  \href{http://dx.doi.org/10.1103/PhysRevLett.111.222301}{{\em Phys. Rev.
  Lett.} {\bfseries 111} (2013) 222301},
\href{http://arxiv.org/abs/1307.5530}{{\ttfamily arXiv:1307.5530 [nucl-ex]}}.

\bibitem{Lee:2007wr}
S.~H. Lee, K.~Ohnishi, S.~Yasui, I.-K. Yoo, and C.-M. Ko, ``{$\Lambda_{\rm c}$
  enhancement from strongly coupled quark-gluon plasma}'',
  \href{http://dx.doi.org/10.1103/PhysRevLett.100.222301}{{\em Phys. Rev.
  Lett.} {\bfseries 100} (2008) 222301},
\href{http://arxiv.org/abs/0709.3637}{{\ttfamily arXiv:0709.3637 [nucl-th]}}.

\bibitem{Kuznetsova:2006bh}
I.~Kuznetsova and J.~Rafelski, ``{Heavy flavor hadrons in statistical
  hadronization of strangeness-rich QGP}'',
  \href{http://dx.doi.org/10.1140/epjc/s10052-007-0268-9}{{\em Eur. Phys. J.}
  {\bfseries C51} (2007) 113--133},
\href{http://arxiv.org/abs/hep-ph/0607203}{{\ttfamily arXiv:hep-ph/0607203
  [hep-ph]}}.

\bibitem{Acharya:2017kfy}
{\bfseries ALICE} Collaboration, S.~Acharya {\em et~al.}, ``{$\Lambda_{\rm
  c}^+$ production in pp collisions at $\sqrt{s} = 7$ TeV and in p-Pb
  collisions at $\sqrt{s_{\rm NN}} = 5.02$ TeV}'',
  \href{http://dx.doi.org/10.1007/JHEP04(2018)108}{{\em JHEP} {\bfseries 04}
  (2018) 108},
\href{http://arxiv.org/abs/1712.09581}{{\ttfamily arXiv:1712.09581 [nucl-ex]}}.

\bibitem{Aaij:2013mga}
{\bfseries LHCb} Collaboration, R.~Aaij {\em et~al.}, ``{Prompt charm
  production in pp collisions at $\sqrt{s}=7$~{TeV}}'',
  \href{http://dx.doi.org/10.1016/j.nuclphysb.2013.02.010}{{\em Nucl. Phys.}
  {\bfseries B871} (2013) 1--20},
\href{http://arxiv.org/abs/1302.2864}{{\ttfamily arXiv:1302.2864 [hep-ex]}}.

\bibitem{Maciula:2018iuh}
R.~Maciuła and A.~Szczurek, ``{Production of $\Lambda_c$ baryons at the LHC
  within the $k_T$-factorization approach and independent parton fragmentation
  picture}'', \href{http://dx.doi.org/10.1103/PhysRevD.98.014016}{{\em Phys.
  Rev.} {\bfseries D98} no.~1, (2018) 014016},
\href{http://arxiv.org/abs/1803.05807}{{\ttfamily arXiv:1803.05807 [hep-ph]}}.

\bibitem{Aaij:2018iyy}
{\bfseries LHCb} Collaboration, R.~Aaij {\em et~al.}, ``{Prompt $\Lambda^+_c$
  production in $p\mathrm{Pb}$ collisions at $\sqrt{s_{NN}} = 5.02$ TeV}'',
\href{http://arxiv.org/abs/1809.01404}{{\ttfamily arXiv:1809.01404 [hep-ex]}}.

\bibitem{Li:2017zuj}
H.-H. Li, F.-L. Shao, J.~Song, and R.-Q. Wang, ``{Production of single-charm
  hadrons by quark combination mechanism in $p$-Pb collisions at
  $\sqrt{s_{NN}}=5.02$ TeV}'',
  \href{http://dx.doi.org/10.1103/PhysRevC.97.064915}{{\em Phys. Rev.}
  {\bfseries C97} no.~6, (2018) 064915},
\href{http://arxiv.org/abs/1712.08921}{{\ttfamily arXiv:1712.08921 [hep-ph]}}.

\bibitem{Song:2018tpv}
J.~Song, H.-h. Li, and F.-l. Shao, ``{New feature of low $p_{T}$ charm quark
  hadronization in $pp$ collisions at $\sqrt{s}=7$ TeV}'',
  \href{http://dx.doi.org/10.1140/epjc/s10052-018-5817-x}{{\em Eur. Phys. J.}
  {\bfseries C78} no.~4, (2018) 344},
\href{http://arxiv.org/abs/1801.09402}{{\ttfamily arXiv:1801.09402 [hep-ph]}}.

\bibitem{Aamodt:2008zz}
{\bfseries ALICE} Collaboration, K.~Aamodt {\em et~al.}, ``{The ALICE
  experiment at the CERN LHC}'',
\href{http://dx.doi.org/10.1088/1748-0221/3/08/S08002}{{\em JINST} {\bfseries
  3} (2008) S08002}.

\bibitem{Acharya:2018hre}
{\bfseries ALICE} Collaboration, S.~Acharya {\em et~al.}, ``{Measurement of
  D$^0$, D$^+$, D$^{*+}$ and D$^+_{\rm s}$ production in Pb-Pb collisions at
  $\sqrt{s_{\rm NN}}= 5.02$ TeV}'', {\em Submitted to: JHEP} (2018) ,
\href{http://arxiv.org/abs/1804.09083}{{\ttfamily arXiv:1804.09083 [nucl-ex]}}.

\bibitem{Tanabashi:2018pdg}
{\bfseries Particle Data Group} Collaboration, M.~Tanabashi {\em et~al.},
  ``{Review of Particle Physics}'',
{\em Phys. Rev.} {\bfseries D98} (2018) 030001.

\bibitem{Abelev:2014ffa}
{\bfseries ALICE} Collaboration, B.~Abelev {\em et~al.}, ``{Performance of the
  ALICE Experiment at the CERN LHC}'',
  \href{http://dx.doi.org/10.1142/S0217751X14300440}{{\em Int. J. Mod. Phys.}
  {\bfseries A29} (2014) 1430044},
\href{http://arxiv.org/abs/1402.4476}{{\ttfamily arXiv:1402.4476 [nucl-ex]}}.

\bibitem{Aamodt:2010aa}
{\bfseries ALICE} Collaboration, K.~Aamodt {\em et~al.}, ``{Alignment of the
  ALICE Inner Tracking System with cosmic-ray tracks}'',
  \href{http://dx.doi.org/10.1088/1748-0221/5/03/P03003}{{\em JINST} {\bfseries
  5} (2010) P03003},
\href{http://arxiv.org/abs/1001.0502}{{\ttfamily arXiv:1001.0502
  [physics.ins-det]}}.

\bibitem{Alme:2010ke}
J.~Alme {\em et~al.}, ``{The ALICE TPC, a large 3-dimensional tracking device
  with fast readout for ultra-high multiplicity events}'',
  \href{http://dx.doi.org/10.1016/j.nima.2010.04.042}{{\em Nucl. Instrum.
  Meth.} {\bfseries A622} (2010) 316--367},
\href{http://arxiv.org/abs/1001.1950}{{\ttfamily arXiv:1001.1950
  [physics.ins-det]}}.

\bibitem{Akindinov:2013tea}
A.~Akindinov {\em et~al.}, ``{Performance of the ALICE Time-Of-Flight detector
  at the LHC}'',
\href{http://dx.doi.org/10.1140/epjp/i2013-13044-x}{{\em Eur. Phys. J. Plus}
  {\bfseries 128} (2013) 44}.

\bibitem{Abbas:2013taa}
{\bfseries ALICE} Collaboration, E.~Abbas {\em et~al.}, ``{Performance of the
  ALICE VZERO system}'',
  \href{http://dx.doi.org/10.1088/1748-0221/8/10/P10016}{{\em JINST} {\bfseries
  8} (2013) P10016},
\href{http://arxiv.org/abs/1306.3130}{{\ttfamily arXiv:1306.3130 [nucl-ex]}}.

\bibitem{ALICE-PUBLIC-2018-011}
{\bfseries ALICE} Collaboration, ``{Centrality determination in heavy ion
  collisions}'', {\em
  {\href{https://cds.cern.ch/record/2636623?ln=en}{ALICE-PUBLIC-2018-011}}}
  (2018) .

\bibitem{Armenteros}
J.~Podolanski and R.~Armenteros, ``{III. Analysis of V-events}'', {\em Phylos.
  Mag.} {\bfseries 45} (1954) 13--30.

\bibitem{Abelev:2014ipa}
{\bfseries ALICE} Collaboration, B.~Abelev {\em et~al.}, ``{Azimuthal
  anisotropy of D meson production in Pb-Pb collisions at $\sqrt{s_{\rm NN}} =
  2.76$ TeV}'', \href{http://dx.doi.org/10.1103/PhysRevC.90.034904}{{\em Phys.
  Rev.} {\bfseries C90} no.~3, (2014) 034904},
\href{http://arxiv.org/abs/1405.2001}{{\ttfamily arXiv:1405.2001 [nucl-ex]}}.

\bibitem{Wang:1991hta}
X.-N. Wang and M.~Gyulassy, ``{HIJING: A Monte Carlo model for multiple jet
  production in pp, pA and AA collisions}'',
\href{http://dx.doi.org/10.1103/PhysRevD.44.3501}{{\em Phys. Rev.} {\bfseries
  D44} (1991) 3501--3516}.

\bibitem{Sjostrand:2006za}
T.~Sjostrand, S.~Mrenna, and P.~Z. Skands, ``{PYTHIA 6.4 Physics and Manual}'',
  \href{http://dx.doi.org/10.1088/1126-6708/2006/05/026}{{\em JHEP} {\bfseries
  05} (2006) 026},
\href{http://arxiv.org/abs/hep-ph/0603175}{{\ttfamily arXiv:hep-ph/0603175
  [hep-ph]}}.

\bibitem{Brun:1994aa}
R.~Brun, F.~Bruyant, F.~Carminati, S.~Giani, M.~Maire, A.~McPherson,
  G.~Patrick, and L.~Urban, ``{GEANT Detector Description and Simulation
  Tool}'', \href{http://dx.doi.org/10.17181/CERN.MUHF.DMJ1}{{\em CERN Program
  Library Long Writeup} (1994) }
CERN-W5013.

\bibitem{Cacciari:1998it}
M.~Cacciari, M.~Greco, and P.~Nason, ``{The $p_{\rm T}$ spectrum in heavy
  flavor hadroproduction}'',
  \href{http://dx.doi.org/10.1088/1126-6708/1998/05/007}{{\em JHEP} {\bfseries
  05} (1998) 007},
\href{http://arxiv.org/abs/hep-ph/9803400}{{\ttfamily arXiv:hep-ph/9803400
  [hep-ph]}}.

\bibitem{Cacciari:2001td}
M.~Cacciari, S.~Frixione, and P.~Nason, ``{The $p_{\rm T}$ spectrum in heavy
  flavor photoproduction}'',
  \href{http://dx.doi.org/10.1088/1126-6708/2001/03/006}{{\em JHEP} {\bfseries
  03} (2001) 006},
\href{http://arxiv.org/abs/hep-ph/0102134}{{\ttfamily arXiv:hep-ph/0102134
  [hep-ph]}}.

\bibitem{ALICE:2011aa}
{\bfseries ALICE} Collaboration, B.~Abelev {\em et~al.}, ``{Measurement of
  charm production at central rapidity in proton-proton collisions at $\sqrt{s}
  = 7$ TeV}'', \href{http://dx.doi.org/10.1007/JHEP01(2012)128}{{\em JHEP}
  {\bfseries 01} (2012) 128},
\href{http://arxiv.org/abs/1111.1553}{{\ttfamily arXiv:1111.1553 [hep-ex]}}.

\bibitem{Gladilin:2014tba}
L.~Gladilin, ``{Fragmentation fractions of $c$ and $b$ quarks into charmed
  hadrons at LEP}'',
  \href{http://dx.doi.org/10.1140/epjc/s10052-014-3250-3}{{\em Eur. Phys. J.}
  {\bfseries C75} no.~1, (2015) 19},
\href{http://arxiv.org/abs/1404.3888}{{\ttfamily arXiv:1404.3888 [hep-ex]}}.

\bibitem{Lange:2001uf}
D.~J. Lange, ``{The EvtGen particle decay simulation package}'',
\href{http://dx.doi.org/10.1016/S0168-9002(01)00089-4}{{\em Nucl. Instrum.
  Meth.} {\bfseries A462} (2001) 152--155}.

\bibitem{Miller:2007ri}
M.~L. Miller, K.~Reygers, S.~J. Sanders, and P.~Steinberg, ``{Glauber modeling
  in high energy nuclear collisions}'',
  \href{http://dx.doi.org/10.1146/annurev.nucl.57.090506.123020}{{\em Ann. Rev.
  Nucl. Part. Sci.} {\bfseries 57} (2007) 205--243},
\href{http://arxiv.org/abs/nucl-ex/0701025}{{\ttfamily arXiv:nucl-ex/0701025
  [nucl-ex]}}.

\bibitem{Loizides:2017ack}
C.~Loizides, J.~Kamin, and D.~d'Enterria, ``{Improved Monte Carlo Glauber
  predictions at present and future nuclear colliders}'',
  \href{http://dx.doi.org/10.1103/PhysRevC.97.054910}{{\em Phys. Rev.}
  {\bfseries C97} no.~5, (2018) 054910},
\href{http://arxiv.org/abs/1710.07098}{{\ttfamily arXiv:1710.07098 [nucl-ex]}}.

\bibitem{Khachatryan:2016ypw}
{\bfseries CMS} Collaboration, V.~Khachatryan {\em et~al.}, ``{Suppression and
  azimuthal anisotropy of prompt and nonprompt ${\mathrm{J}}/\psi $ production
  in PbPb collisions at $\sqrt{{s_{_{\text {NN}}}}} =2.76$ $\,\mathrm{TeV}$}'',
  \href{http://dx.doi.org/10.1140/epjc/s10052-017-4781-1}{{\em Eur. Phys. J.}
  {\bfseries C77} no.~4, (2017) 252},
\href{http://arxiv.org/abs/1610.00613}{{\ttfamily arXiv:1610.00613 [nucl-ex]}}.

\bibitem{Sirunyan:2017xss}
{\bfseries CMS} Collaboration, A.~M. Sirunyan {\em et~al.}, ``{Nuclear
  modification factor of D$^0$ mesons in PbPb collisions at
  $\sqrt{s_\mathrm{NN}} = 5.02$ TeV}'',
  \href{http://dx.doi.org/10.1016/j.physletb.2018.05.074}{{\em Phys. Lett.}
  {\bfseries B782} (2018) 474--496},
\href{http://arxiv.org/abs/1708.04962}{{\ttfamily arXiv:1708.04962 [nucl-ex]}}.

\bibitem{Das:2016llg}
S.~K. Das, J.~M. Torres-Rincon, L.~Tolos, V.~Minissale, F.~Scardina, and
  V.~Greco, ``{Propagation of heavy baryons in heavy-ion collisions}'',
  \href{http://dx.doi.org/10.1103/PhysRevD.94.114039}{{\em Phys. Rev.}
  {\bfseries D94} no.~11, (2016) 114039},
\href{http://arxiv.org/abs/1604.05666}{{\ttfamily arXiv:1604.05666 [nucl-th]}}.

\bibitem{MartinezGarcia:2007hf}
G.~Martinez-Garcia, S.~Gadrat, and P.~Crochet, ``{Consequences of a
  $\Lambda_{\rm c}/{\rm D}$ enhancement effect on the non-photonic electron
  nuclear modification factor in central heavy ion collisions at RHIC
  energy}'', \href{http://dx.doi.org/10.1016/j.physletb.2008.07.061,
  10.1016/j.physletb.2008.01.079}{{\em Phys. Lett.} {\bfseries B663} (2008)
  55--60}, \href{http://arxiv.org/abs/0710.2152}{{\ttfamily arXiv:0710.2152
  [hep-ph]}}.
[Erratum: Phys. Lett. {\bf B666}, (2008) 533].

\bibitem{Acharya:2018qsh}
{\bfseries ALICE} Collaboration, S.~Acharya {\em et~al.}, ``{Transverse
  momentum spectra and nuclear modification factors of charged particles in pp,
  p-Pb and Pb-Pb collisions at the LHC}'',
\href{http://arxiv.org/abs/1802.09145}{{\ttfamily arXiv:1802.09145 [nucl-ex]}}.

\bibitem{Xie:2017jcq}
{\bfseries STAR} Collaboration, G.~Xie, ``{$\Lambda_{c}$ Production in Au+Au
  Collisions at $\sqrt{s_{NN}}$ = 200 GeV measured by the STAR experiment}'',
  \href{http://dx.doi.org/10.1016/j.nuclphysa.2017.06.004}{{\em Nucl. Phys.}
  {\bfseries A967} (2017) 928--931},
\href{http://arxiv.org/abs/1704.04353}{{\ttfamily arXiv:1704.04353 [nucl-ex]}}.

\bibitem{Acharya:2017jgo}
{\bfseries ALICE} Collaboration, S.~Acharya {\em et~al.}, ``{Measurement of
  D-meson production at mid-rapidity in pp collisions at ${\sqrt{s}=7}$ TeV}'',
  \href{http://dx.doi.org/10.1140/epjc/s10052-017-5090-4}{{\em Eur. Phys. J.}
  {\bfseries C77} no.~8, (2017) 550},
\href{http://arxiv.org/abs/1702.00766}{{\ttfamily arXiv:1702.00766 [hep-ex]}}.

\bibitem{AliceUpgrade}
{\bfseries ALICE} Collaboration, B.~Abelev {\em et~al.}, ``{Upgrade of the
  ALICE Experiment: Letter Of Intent}'',
\href{http://dx.doi.org/10.1088/0954-3899/41/8/087001}{{\em J. Phys.}
  {\bfseries G41} (2014) 087001}.

\bibitem{Abelev:2014dna}
{\bfseries ALICE} Collaboration, B.~Abelev {\em et~al.}, ``{Technical Design
  Report for the Upgrade of the ALICE Inner Tracking System}'',
\href{http://dx.doi.org/10.1088/0954-3899/41/8/087002}{{\em J. Phys.}
  {\bfseries G41} (2014) 087002}.

\end{thebibliography}\endgroup


\providecommand{\href}[2]{#2}\begingroup\raggedright\endgroup
